\documentclass[fleqn,twoside]{article}
\usepackage{espcrc2}

\usepackage{graphicx}
\usepackage[figuresright]{rotating}

\newcommand{\AmS}{{\protect\the\textfont2
  A\kern-.1667em\lower.5ex\hbox{M}\kern-.125emS}}

\title{Neutrino Oscillation Search at MiniBooNE}

\author{Z. Djurcic\thanks{{\it Email address:} zdjurcic@nevis.columbia.edu} \hspace{2mm} 
for the MiniBooNE Collaboration
\address[MCSD]{Columbia University, New York, NY 10027, USA}%
        \thanks{MiniBooNE Experiment is supported by NSF and DOE.}}
       
\begin{document}

\begin{abstract}
This article reports the status of a $\nu_{\mu} \rightarrow \nu_{e}$ 
oscillation search in MiniBooNE (Booster Neutrino Experiment) experiment.
If an appearance signal is observed, it will imply Physics Beyond 
the Standard Model such as the existence of light sterile neutrino.
\vspace{1pc}
\end{abstract}

\maketitle

\section{INTRODUCTION}

In recent years, the solar-neutrino~\cite{solarE}, reactor-neutrino~\cite{reactorE}, 
atmospheric-neutrino~\cite{atmE}, and accelerator-neutrino~\cite{accE} experiments 
have confirmed the existence of neutrino oscillations.
These results implied the existence of two independent $\Delta m^2$ regions,
with  $\Delta m^2 \sim 8 \times 10^{-5} eV^2$ in the solar, and
with $\Delta m^2 \sim 3 \times 10^{-3} eV^2$ in the atmospheric sector.
The Standard Model incorporates  two separate oscillation regions with three known
neutrino flavors: $\nu_{e}$, $\nu_{\mu}$, and $\nu_{\tau}$.
However, an unconfirmed evidence for neutrino oscillations came from the LSND~\cite{lsndE}
experiment with $\Delta m^2$ at $\sim 1 eV^2$ value.
The discovery of nonzero neutrino masses through the neutrino oscillations
has raised a number of very interesting questions about the neutrinos and 
their connections to other areas of physics and astrophysics.
One question is whether there are sterile neutrinos that do not 
participate in the standard weak interactions. 
This question is primarily being addressed by the MiniBooNE 
experiment.
The MiniBooNE experiment will confirm or refute the LSND result
with higher statistics and different sources of systematic error.
If the LSND neutrino oscillation evidence is confirmed, it will, together
with solar, reactor, atmospheric and accelerator oscillation
data, imply Physics Beyond the Standard Model such as the existence
of light sterile neutrino~\cite{sorel}.
While LSND observed an excess of $\bar\nu_{e}$ events in a 
$\bar\nu_{\mu}$ beam, MiniBooNE is a $\nu_{\mu} \rightarrow \nu_{e}$ 
search.

\section{THE EXPERIMENT}
MiniBooNE is a fixed target experiment currently taking data at Fermi National 
Accelerator Laboratory. The neutrino beam is produced from 8.89 $GeV/c$ protons 
impinging on a 71 $cm$ long and 1 $cm$ diameter beryllium target. The target is located 
inside a magnetic focusing horn that increases the neutrino flux at the detector 
by a factor of $\sim$5, and can operate in both negative and positive polarities 
for $\nu$ and $\bar{\nu}$ running.  
MiniBooNE collected approximately $6 \times 10^{20}$ protons on target (POT)
in neutrino mode. This data sample is currently used in
the neutrino oscillation analysis. Mesons
produced in the target decay-in-flight in a 50 $m$ long decay pipe.
The neutrino beam is composed of $\nu_{\mu}$ from
$K^{+}/\pi^{+} \rightarrow \mu^{+} + \nu_{\mu}$
decays.  The neutrino beam propagates through a 450 $m$ of a dirt 
absorber before entering the detector.
There is a small contamination from $\nu_{e}$.
The processes that contribute to the intrinsic $\nu_{e}$
in the beam are $\mu^{+} \rightarrow e^{+} \nu_{e} \bar{\nu}_{\mu}$,
$K^{+} \rightarrow \pi^{0} e^{+} \nu_{e}$, 
and $K_{L}^{0} \rightarrow \pi^{\pm} e^{\pm} \nu_{e}$.
Early in 2006, MiniBooNE switched
the polarity of the horn to select negative sign mesons.

\begin{figure}[htb]
\vspace{9pt}
\resizebox{.9\columnwidth}{!}
   {\includegraphics[width=14.cm, height=13.cm]{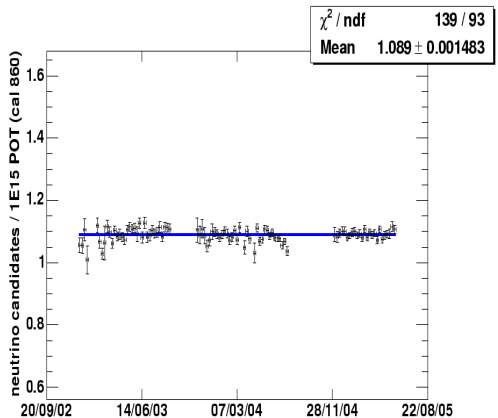}}
\caption{Calibrated neutrino rate per proton on target.}
\label{fig:Events_per_POT}
\end{figure}

The MiniBooNE detector is a 12.2 $m$ diameter sphere filled with 800 tons of
pure mineral oil.  The center of the MiniBooNE neutrino detector is
positioned $L = 541 m$ from the front of the beryllium target.
The vessel consists of two optically isolated
regions divided by a support structure located at 5.5 $m$ radius.  The inner
volume is the neutrino target region, while the outer volume forms the veto
region.  There are 1280 8-inch photo-tubes (PMTs) pointed inward 
providing 10\% coverage, and 292 outward-faced PMTs in the veto region.
Data analysis require a fiducial volume cut at 500 $cm$ from the center 
of the tank to ensure good event reconstruction, resulting in 
a 445 ton target region. The outer volume serves as a veto shield for
identifying particles both entering and leaving the detector.
The rate of neutrino candidates per proton delivered to the
target was constant ($1.089\times10^{-15}$ $\nu$/POT) 
over the period of data acquisition, as demonstrated in 
Figure~\ref{fig:Events_per_POT}.
Protons delivered to the MiniBooNE target 
are monitored by two toroids located in the beam line. 
The main MiniBooNE trigger is an accelerator signal indicating
a beam spill. Every beam trigger opens a 19.2 $\mu s$ window
in which all events are recorded. Other triggers include
a supernova trigger, a random strobe trigger for beam-off
measurements, a laser calibration trigger, cosmic muon 
triggers, and a trigger to record neutrino events from the
NuMI beam line.  

The interaction point, event time, energies, and the 
particle tracks are recorded from the times and charges of the PMT hits
in the detector.  
Neutrino induced events are identified by requiring that the event
is observed during the beam spill, have fewer than 6 veto PMT hits,
and greater than 200 tank PMT hits. These simple cuts yield a cosmic ray
rejection that is better than 1000:1.

The MiniBooNE detector is calibrated using a laser system, cosmic 
rays and neutrino interactions. The calibration procedures cover 
the full energy range from 50 to 1000 $MeV$ using different event 
types. Stopping cosmic ray muons are used to calibrate the energy 
scale for muon-type events and measure the position and angle 
reconstruction resolution, when their path length can be identified. 
This is accomplished with a scintillator hodoscope on the top of the 
detector, combined with scintillation cubes at various positions within 
the detector volume. Analysis of the cosmic muons indicates that 
the energy and angle resolution of muons is 10\% and $6.4^{o}$, 
respectively.
Observed Michel electrons from muon decay are used to calibrate 
the energy scale of electron-type events at the 52.8 $MeV$ Michel 
endpoint, with 12\% resolution. Reconstructed $\pi^{0}$ events 
provide another electron-like calibration source. The 
photons that are emitted in $\pi^{0}$ decays span a
considerable range to over 1000 $MeV$. The $\pi^{0}$ mass
derived from reconstructed energies and directions
of two $\gamma$-rays has a peak at 136.3$\pm$0.8 $MeV$.
The resolution on the reconstructed $\pi^0$ mass peak
is 20 $MeV$. This is in an excellent agreement with the 135.0 $MeV$ 
expectation, providing a check on the energy scale and 
the reconstruction over the full energy range of interest for 
the $\nu_{e}$ appearance analysis.

\section{NEUTRINO FLUX, CROSS SECTION, AND DETECTOR MODELING}

The flux modeling uses a Geant4-based simulation of beam line geometry.
Hadron production in the target is based on Sanford-Wang parametrization
of $p-Be$ cross-section, with parameters determined by a global fit
to $p-Be$ particle production data. Simulated neutrino flux has
an energy distribution with a peak at $\sim 0.7 GeV$.
Therefore, the average $L/E_{\nu}$ ratio is $\sim 0.8 km/GeV$
compared to LSND's $L/E_{\nu} \sim 1 km/GeV$.

The NUANCEv3~\cite{nuance} event-generator simulates
interactions in the detector. The cross section model describes
the various neutrino interaction processes on $CH_{2}$, which
include the Llewellyn-Smith free nucleon quasi-elastic cross section,
the Rein-Seghal resonant and coherent pion production cross 
section, a Smith-Moniz Fermi gas model, and final state
interactions based on $\pi$-Carbon scattering data.

The MiniBooNE detector is modeled using an extended 
GEANT3-based simulation.  An extended light
propagation model of the detector describes the emission of
optical and near-UV photons via Cerenkov radiation and
scintillation. Each photon is individually 
tracked, undergoing scattering, fluorescence, and reflection,
until it is absorbed.  The response of the electronics to the
photoelectrons resulting from photons hitting PMT's 
is simulated, with the final output being digitized waveforms
simulating the charge and time channels of the electronics in
a form identical to that used for data.

\section{THE BACKGROUNDS IN THE APPEARANCE SIGNAL}

The signature of an oscillation event is the $\nu_{e} + n \rightarrow e + p$
reaction.
The backgrounds in the oscillation analysis are divided in two main
categories: intrinsic $\nu_{e}$ events in the beam, and $\nu_{\mu}$
events that are mis-identified as $\nu_{e}$ events. 

The beam that arrives at the detector is almost pure $\nu_{\mu}$ with a 
small (0.6\%) contamination of $\nu_{e}$ coming from muon and kaon decays 
in the decay pipe. 
The $\nu_{e}$ from $\mu$-decay
are directly tied to the observed $\nu_{\mu}$ interactions. Taking
into account a small solid angle subtended by the MiniBooNE detector,
the pion energy distribution can be determined from the energy of
the observed $\nu_{\mu}$ events. The pion energy spectrum is then
used to predict the $\nu_{e}$ from $\mu$-decay.
A second source of $\nu_{e}$ originates from $K_{e3}$-decay. This
component is constrained using high energy $\nu_{\mu}$
charged current quasi-elastic (CCQE) events, that originate primarily 
from kaon decays.

\begin{figure}[htb]
\vspace{9pt}
\resizebox{.9\columnwidth}{!}
   {\includegraphics[width=14.cm, height=16.cm]{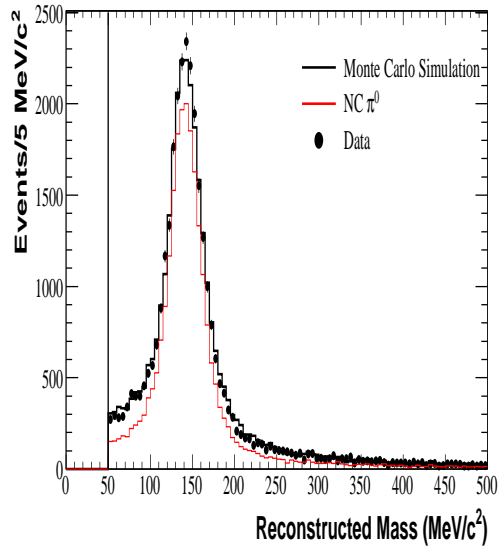}}
\caption{(Preliminary) Mass distribution of $\pi^{0}$ candidates obtained 
using a two ring likelihood fit. The data are the black points, 
the black line is Monte Carlo simulated neutrino interactions, 
while the red line is the subset of these events 
which have at least one $pi^0$ in the final state.}
\label{fig:pi0_mass}
\end{figure}

Mis-identified $\nu_{\mu}$ events are $\pi^{0}$, $\Delta$-decays, and 
$\nu_{\mu}$ CCQE events.
Most $\pi^{0}$ are identified by the reconstruction of two Cerenkov 
rings produced by two decay $\gamma$-rays, as shown in Figure~\ref{fig:pi0_mass}.
However, the decay of $\pi^{0}$ 
can appear much like primary electron emerging from a $\nu_{e}$
charged current interaction if one of the gammas from the decay 
overlaps the other, or is too low in energy to be detected.
Over 99\% of the NC $\pi^{0}$ are rejected in the appearance analysis.
In addition to its primary
decay $\Delta \rightarrow \pi\;N$, the $\Delta$ resonance has a branching
fraction of 0.56\% to the $\gamma \; N$ final state. The $\gamma$-ray may mimic an
electron from $\nu_{e}$ interaction. The rate of $\Delta$ production in neutral
current interactions can be estimated from the data, using the sample of
reconstructed $\pi^{0}$ decays. 
Most $\nu_{\mu}$ events can be
easily identified by their penetration into the veto region when exiting muons
fire the veto, or by muons stopping in the inner detector and producing a Michel 
electron after a few microseconds.

\section{SIGNAL SEPARATION FROM THE BACKGROUND}

\begin{figure}[htb]
\vspace{9pt}
\resizebox{.9\columnwidth}{!}
   {\includegraphics[width=14.cm, height=16.cm]{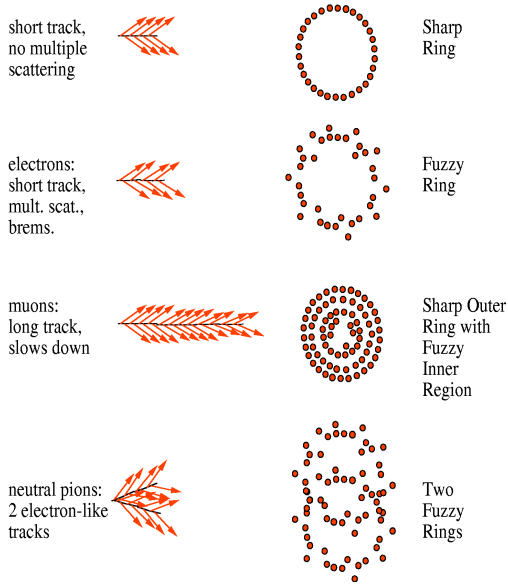}}
\caption{Cartoon showing how particle identification acquire an input from the Cerenkov 
rings. From the top to the bottom, the rings are: an ideal particle, an electron, 
a stopping muon, and $\pi^0$.}
\label{fig:PID}
\end{figure}

\begin{figure}[htb]
\vspace{9pt}
\resizebox{.9\columnwidth}{!}
   {\includegraphics[width=12.cm, height=10.cm]{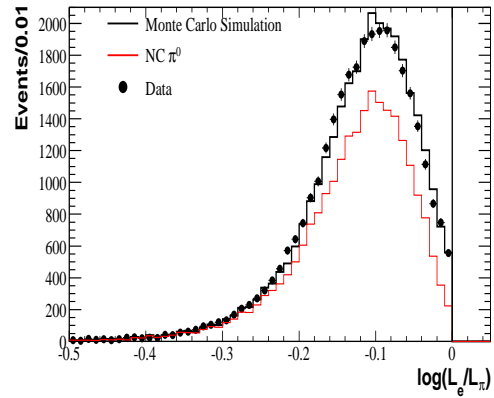}}
\caption{(Preliminary) The logarithm of the likelihood ratio 
formed from fitting neutral current $\pi^{0}$ candidates under 
a single electron ring hypothesis and a two ring hypothesis.
The data are the black points, 
the black line is Monte Carlo simulated neutrino interactions, while the 
red line is the subset of these events which have at least one $\pi^{0}$ 
in the final state.}
\label{fig:logL}
\end{figure}

Particle identification (PID) is performed by different algorithms 
that use the difference in characteristics of the Cerenkov and 
scintillation light associated with electrons, muons, protons or $\pi^{0}$'s,
as shown in Figure~\ref{fig:PID}.
These algorithms include a maximum likelihood method
and boosted decision trees~\cite{pid_papers}. 
Figure~\ref{fig:logL} shows the logarithm of the likelihood ratio 
formed from fitting neutral current $\pi^{0}$ candidates under 
a single electron ring hypothesis and a two ring hypothesis where 
the kinematics are fixed to give the nominal $\pi^{0}$ mass. Electron-like 
events should have larger positive values, while $\pi^{0}$'s should have 
negative values. The events are selected by requiring no decay electrons 
following the primary interaction, and requiring the veto to have less 
than six hits to ensure there is no cosmic muon contamination 
and the tank to have greater than 200 hits to suppress decay 
electrons from cosmic muons. The events are then fit under the 
single electron and muon ring hypotheses and a ratio is formed with the 
resulting likelihoods. Events with likelihood ratios favoring the muon 
hypothesis are rejected. The event is then fit with two ring fits, 
both with the mass free and fixed to the nominal $\pi^{0}$ mass. 
In the case from Figure~\ref{fig:PID}, separation between $\pi^{0}$ and $\nu_{e}$
candidates is achieved. Similar PID separation is performed 
between electrons and other types of detected events.
An example of a boosted decision
tree where the muon/electron separation was measured with cosmic ray
muons and associated electrons is given in \cite{djurcic_panic05}. 
The PID removes $\sim 99.9\%$ of $\nu_{\mu}$ CCQE interactions,
$\sim 99\%$ effective $\pi^{0}$ producing interactions, and 
preserves a high efficiency for $\nu_{e}$ interactions.
An optimized PID is expected to allow a small contamination of 
an oscillation signal with mis-identified  $\pi^{0}$ ($\sim 83 \%$),
$\Delta$-decays ($\sim 7 \%$) , and $\nu_{\mu}$ CCQE events ($\sim 10 \%$).

\section{BLIND ANALYSIS STRATEGY AND CROSS-CHECKS}

MiniBooNE is conducting a blind analysis in order to complete an unbiased 
oscillation search. That means that the region where the oscillation
$\nu_{e}$ candidates are expected is closed for the analysis.
In the example given in Figure~\ref{fig:logL}, the mass from the fit 
is required to be greater than 50 $MeV/c^2$. The likelihood ratio 
of the single electron fit and the fixed mass 
two ring fit is required to favor the $\pi^{0}$ hypothesis. 
The latter two requirements are for blindness, in order to keep
the region with potential oscillation
$\nu_{e}$ candidates out of the analysis reach
until the final set of selection cuts is formed and the PID is optimized.
In practice, the MiniBooNE analysis cannot use its own data to 
verify PID algorithms with an oscillation-like $\nu_{e}$ data sample. 
However, an important cross-check of electron event reconstruction and particle
identification comes from NuMI events observed in the MiniBooNE detector.
Neutrinos produced by the decay of mesons moving along the NuMI beamline 
and in the vicinity of the NuMI target may reach the MiniBooNE detector.
NuMI events consist of $\nu_{e}$, $\nu_{\mu}$, $\pi^{\pm}$, $\pi^{0}$ and $\Delta$
over the range of energies relevant to the appearance analysis. 
A kaon and non-kaon fractions in NuMI Monte Carlo are extracted from a fit
to the data. Such Monte Carlo is then compared to our Monte Carlo
prediction. The NuMI sample therefore allows an independent check
of the MiniBooNE PID algorithms performance in isolating $\nu_{e}$.
Other important checks are performed with either MiniBooNE or external data.
A complete list of cross-checks is given in Table~\ref{tab:crosscheck}.
\begin{table}[htb]
\caption{The MiniBooNE analysis is verified by different experimental cross-checks 
for each event class relevant to $\nu_{e}$ appearance search. }
\label{tab:crosscheck}
\newcommand{\m}{\hphantom{$-$}}
\newcommand{\cc}[1]{\multicolumn{1}{c}{#1}}
\renewcommand{\tabcolsep}{1pc} 
\renewcommand{\arraystretch}{1.2} 
\begin{tabular}{@{}ll}
\hline
$K^{+}$               & HARP~\cite{harp}, External Data\\
                      & MiniBooNE Data                 \\  
$K^{0}$               & E910~\cite{e910}, External Data\\
                      & MiniBooNE Data                 \\
$\mu$                 & MiniBooNE Data                 \\
$\pi^{0}$             & NuMI, MiniBooNE Data           \\
Other ($\Delta$, etc) & NuMI, MiniBooNE Data           \\ 
\hline
\end{tabular}
\end{table}

\section{REMAINING STEPS IN THE OSCILLATION SEARCH}

When the final set of the analysis cuts is determined
and associated systematics evaluated,
the data sample that potentially contains the
oscillation candidates will be un-blinded.
The composition of the final sample will be
predicted by the MiniBooNE Monte Carlo simulation,
with Monte Carlo sample filtered through the same
set of PID cuts.
If MiniBooNE confirms the LSND result, an excess of
events in the data distributions when compared to Monte Carlo
will be observed.
The final event sample will be evaluated using
a $\chi ^2$ function 
\begin{equation}
\label{eq:chi2}
{\small
\chi^2 = \sum_{i=1} (O_i - P_i) (C_{ij})^{-1} (O_j - P_j),
}
\end{equation}
applied to the data and Monte Carlo energy distribution of 
the oscillation candidates.
$O_i$ is the number of observed events in
an energy bin $i$. $P_i$ is the Monte Carlo
prediction that takes into account
oscillation parameters ($\sin^2 2\theta, \Delta m^2$).
The covariance matrix $C_{ij}$ account for the statistical
and systematic uncertainties in the energy bins.
Systematic errors are associated with neutrino flux,
neutrino cross sections, and the detector model.

The flux prediction has the uncertainties corresponding to the production
of $\pi$, $K$, and $K_L$ particles in the MiniBooNE target.
These uncertainties are quantified by a fit to external data sets
from previous experiments on meson production.
The cross section uncertainties are evaluated by continuously 
varying underlying cross section model parameters in the Monte Carlo 
constrained by MiniBooNE data.
Uncertainties on the parameters modeling the optical properties
of the oil in the MiniBooNE detector are constrained by a fit
to the calibration sample of Michel electrons.
The uncertainties are currently being evaluated.

\section{CONCLUSION}
The MiniBooNE experiment will confirm or refute the LSND oscillation signal with
approximately $6 \times 10^{20}$ protons on target collected for the analysis. 
The oscillation appearance analysis is underway, with current work on 
the systematic error evaluation that combines the errors from the $\nu$ flux,
$\nu$ cross-sections, and the detector modeling.

The MiniBooNE Collaboration: 
A.A.~Aguilar-Arevalo, A.O.~Bazarko, S.~J. Brice, B.C.~Brown, L.~Bugel, J.~Cao, L.~Coney, 
 J.M.~Conrad, D.C.~Cox, A.~Curioni, Z.~Djurcic, D.A.~Finley,  B.T.~Fleming, R.~Ford, F.~Garcia,
 G.T.~Garvey, A.~Green, C.~Green, T.L.~Hart, E.~Hawker, R.~Imlay, R.A.~Johnson, 
 P.~Kasper, T.~Katori, T.~Kobilarcik, I.~Kourbanis, S.~Koutsoliotas, J.M.~Link, Y.~Liu, Y.~Liu,
 W.C.~Louis, K.B.M.~Mahn,  W.~Marsh, P.~Martin,  G.~McGregor,  W.~Metcalf, P.D.~Meyers, F.~Mills, G.B.~Mills  
 J.~Monroe, C.D.~Moore, R.H.~Nelson, P.~Nienaber, S.~Ouedraogo, R.B.~Patterson, D.~Perevalov, C.C.~Polly,
 E.~Prebys, J.L.~Raaf, H.~Ray, B.P.~Roe, A.D.~Russell, V.~Sandberg, R.~Schirato, D.~Schmitz, 
 M.H.~Shaevitz, F.C.~Shoemaker, D.~Smith, M.~Sorel, P.~Spentzouris, I.~Stancu, 
 R.J.~Stefanski, M.~Sung, H.A.~Tanaka, R.~Tayloe, M.~Tzanov, M.O.~Wascko, R.~Van de Water 
 D.H.~White, M.J.~Wilking, H.J.~Yang,  G.P.~Zeller, E.D.~Zimmerman.

\end{document}